\documentclass[sigconf, 10pt]{acmart}
\AtBeginDocument{%
  \providecommand\BibTeX{{%
    \normalfont B\kern-0.5em{\scshape i\kern-0.25em b}\kern-0.8em\TeX}}}

\ccsdesc[500]{Human-centered computing~Ubiquitous and mobile devices}
\ccsdesc[300]{Hardware~Power and energy}

\usepackage{subfloat}
\usepackage{ulem}

\usepackage{color,soul}
\usepackage{booktabs} 
\usepackage{subfig} 
\usepackage{graphicx}

\usepackage[utf8]{inputenc}
\usepackage[T1]{fontenc}

\usepackage{siunitx}
\newcommand{\hide}[1]{}

\sethlcolor{yellow}
\ifdefined\scifihidecomments
    \newcommand{\cheng}[1] {}
    \newcommand{\hyunc}[1] {} 
    \newcommand{\yax}[1] {} 
    \newcommand{\tuoc}[1] {} 
    \newcommand{\sony}[1] {} 
    \newcommand{\ReviewerFeedback}[1] {}  
    \newcommand{\fran}[1] {}
    \newcommand{\rz}[1]{}
    \newcommand{\lw}[1] {}
    \newcommand{\mose}[1] {} 
    \newcommand{\ke}[1] {} 

\else
    \definecolor{burntorange}{rgb}{0.8, 0.33, 0.0}
    \definecolor{cadmiumgreen}{rgb}{0.0, 0.42, 0.24}
    \definecolor{cobalt}{rgb}{0.0, 0.28, 0.67}
    \definecolor{amber}{rgb}{1.0, 0.75, 0.0}
    \definecolor{fashionfuchsia}{rgb}{0.96, 0.0, 0.63}
    \definecolor{brightcerulean}{rgb}{0.11, 0.67, 0.84}
    \definecolor{frenchblue}{rgb}{0.0, 0.45, 0.73}
    \definecolor{darkslateblue}{rgb}{0.28, 0.24, 0.55}
    \definecolor{cerulean}{rgb}{0.0, 0.48, 0.65}
    \definecolor{darkpastelgreen}{rgb}{0.01, 0.75, 0.24}

    \newcommand{\hyunc}[1] { \textcolor{burntorange}{[{\hl{hyunc:}} {#1}}]}
    \newcommand{\yax}[1] { \textcolor{magenta}{[{\hl{yaxuan:}} {#1}]}}
    \newcommand{\tuoc}[1] { \textcolor{darkpastelgreen}{[{\hl{tuochao:}} {#1}]}}
    \newcommand{\sony}[1] { \textcolor{blue}{[{\hl{songyun:}} {#1}]}}
    \newcommand{\ReviewerFeedback}[1] { \textcolor{brightcerulean}{[{Reviewer Feedback:} {#1}}]}
    \newcommand{\fran}[1]{\textcolor{burntorange}{[{francois:}{#1}}]}
    \newcommand{\rz}[1]{\textcolor{teal}{[{Ruidong: }{#1}]}}
    \newcommand{\lw}[1]{\textcolor{fashionfuchsia}{[{liuwei:}{#1}}]}
    \newcommand{\mose}[1]{\textcolor{burntorange}{[{mose:}{#1}}]}
    \newcommand{\ke}[1] { \textcolor{red!55!yellow}{[{Ke:} {#1}}]}

\fi

\ifdefined\scifiblind
    \newcommand{\blind}[1]{[omitted for blind review]}
\else
    \newcommand{\blind}[1]{#1} 
\fi

\usepackage{gensymb}
\usepackage{glossaries}
\usepackage{multirow}
\usepackage{pifont}
\usepackage{hyperref}
\usepackage{lipsum}
\usepackage{caption}
\usepackage{siunitx}
\newcommand{\cmark}{\text{\ding{51}}}
\newcommand{\xmark}{\text{\ding{55}}}

\makeglossaries
\newglossaryentry{name}
{
        name=GazeTrak,
        description={Name of the device}
}

\copyrightyear{2024}
\acmYear{2024}
\setcopyright{acmlicensed}\acmConference[ACM MobiCom '24]{The 30th Annual International Conference on Mobile Computing and Networking}{September 30-October 4, 2024}{Washington D.C., DC, USA}
\acmBooktitle{The 30th Annual International Conference on Mobile Computing and Networking (ACM MobiCom '24), September 30-October 4, 2024, Washington D.C., DC, USA}
\acmDOI{10.1145/3636534.3649376}
\acmISBN{979-8-4007-0489-5/24/09}

\begin{document}

\title{\gls{name}: Exploring Acoustic-based Eye Tracking on a Glass Frame}

\author{Ke Li}
\affiliation{%
  \institution{Cornell University}
  \city{Ithaca}
  \country{USA}}
\email{kl975@cornell.edu}
\orcid{0000-0002-4208-7904}

\author{Ruidong Zhang}
\affiliation{%
  \institution{Cornell University}
  \city{Ithaca}
  \country{USA}}
\email{rz379@cornell.edu}
\orcid{0000-0001-8329-0522}

\author{Boao Chen}
\affiliation{%
  \institution{Cornell University}
  \city{Ithaca}
  \country{USA}}
\email{bc526@cornell.edu}
\orcid{0000-0002-3527-9481}

\author{Siyuan Chen}
\affiliation{%
  \institution{Cornell University}
  \city{Ithaca}
  \country{USA}}
\email{sc2489@cornell.edu}
\orcid{0009-0008-4828-3392}

\author{Sicheng Yin}
\affiliation{%
  \institution{University of Edinburgh}
  \city{Edinburgh}
  \country{United Kingdom}}
\email{yinsicheng1999@outlook.com}
\orcid{0000-0002-0165-9750}

\author{Saif Mahmud}
\affiliation{%
  \institution{Cornell University}
  \city{Ithaca}
  \country{USA}}
\email{sm2446@cornell.edu}
\orcid{0000-0002-5283-0765}

\author{Qikang Liang}
\affiliation{%
  \institution{Cornell University}
  \city{Ithaca}
  \country{USA}}
\email{ql75@cornell.edu}
\orcid{0009-0001-9301-625X}

\author{François Guimbretière}
\affiliation{%
  \institution{Cornell University}
  \city{Ithaca}
  \country{USA}}
\email{fvg3@cornell.edu}
\orcid{0000-0002-5510-6799}

\author{Cheng Zhang}
\affiliation{%
  \institution{Cornell University}
  \city{Ithaca}
  \country{USA}}
\email{chengzhang@cornell.edu}
\orcid{0000-0002-5079-5927}

\renewcommand{\shortauthors}{Li et al.}


\begin{abstract}

In this paper, we present \gls{name}, the first acoustic-based eye tracking system on glasses. Our system only needs one speaker and four microphones attached to each side of the glasses. These acoustic sensors capture the formations of the eyeballs and the surrounding areas by emitting encoded inaudible sound towards eyeballs and receiving the reflected signals. These reflected signals are further processed to calculate the echo profiles, which are fed to a customized deep learning pipeline to continuously infer the gaze position. In a user study with 20 participants, \gls{name} achieves an accuracy of $3.6\degree$ within the same remounting session and $4.9\degree$ across different sessions with a refreshing rate of 83.3 Hz and a power signature of 287.9 mW. Furthermore, we report the performance of our gaze tracking system fully implemented on an MCU with a low-power CNN accelerator (MAX78002). In this configuration, the system runs at up to 83.3 Hz and has a total power signature of 95.4 mW with a 30 Hz FPS.

\end{abstract}

\keywords{Eye Tracking, Acoustic Sensing, Smart Glasses, Low-power}


\maketitle

\section{Introduction}
\label{Sec: Introduction}

\begin{figure*}[t]
  \includegraphics[width=0.8 \textwidth]{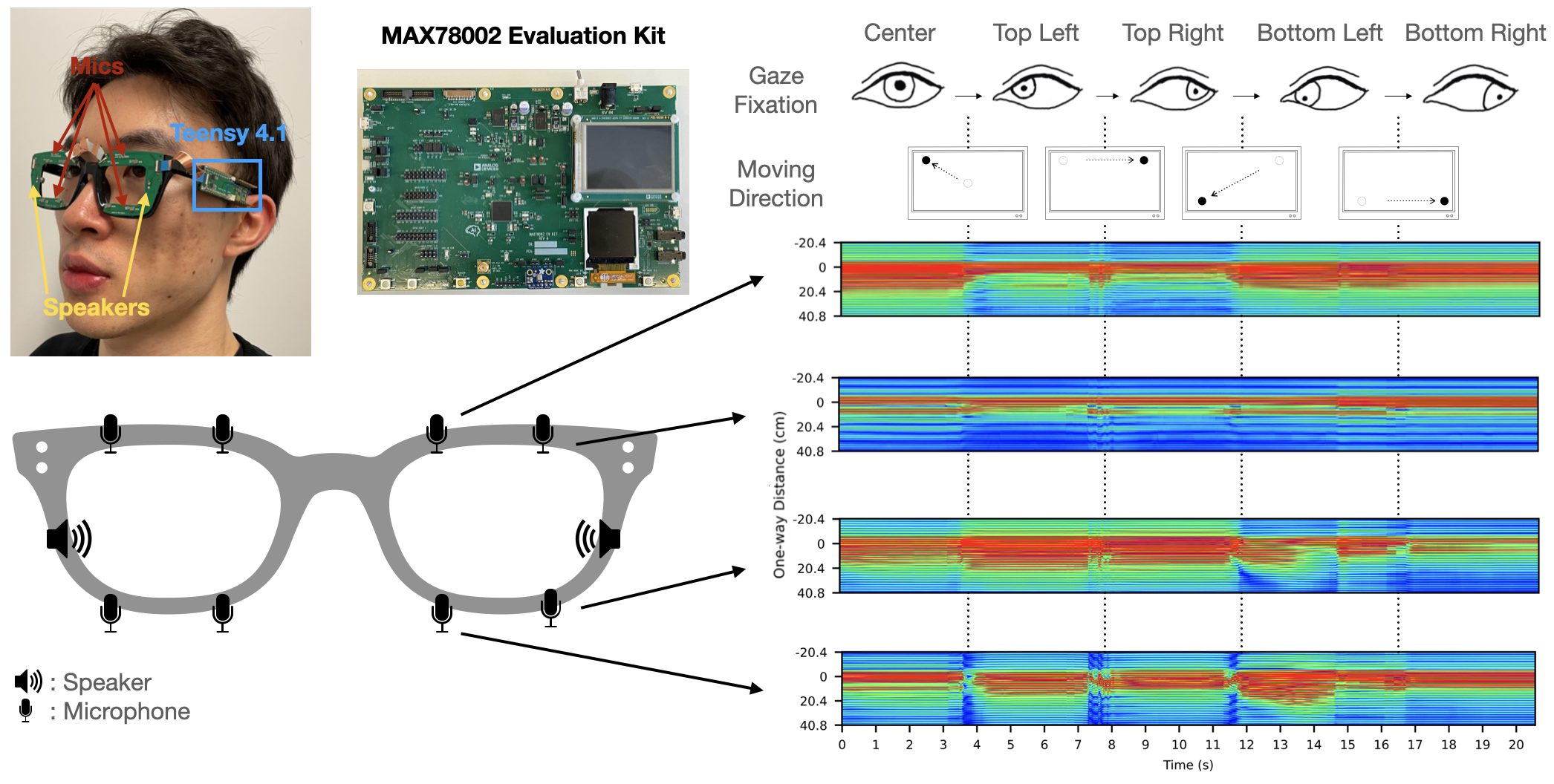}
  \caption{Echo Profiles of Different Microphones when Moving Gaze to Different Regions of The Screen.}
  \label{Fig: patterns}
\end{figure*}

Currently, state-of-the-art eye tracking technologies utilize cameras to capture gaze points. However, cameras-based eye tracking solutions are known to have a relatively high power signature, which may not work well for smart glasses with a relatively small battery capacity. For instance, Tobii Pro Glass 3 \cite{tobiiglasses}, which is considered as one of the best eye tracking glasses, can only last for 1.75 hours with an extended battery capacity of 3400 mAh. When using the battery of a Google Glass (570 mAh), this eye tracking system can only last 18 minutes. 
The limited tracking time has hindered its ability to collect gaze point data in everyday life, which can be highly informative for many applications, such as, monitoring users' mental or physical health conditions \cite{piccardi2007wearcam,vidal2012wearable}, gaze-based input, and attention and interest analysis \cite{dong2022gazby}.

To overcome this challenge, we introduce \gls{name}, which explores utilizing acoustic sensing (known for relatively low power, lightweight, and affordable) to continuously track gaze points on a glass frame. Its sensing principle is based on the fact that eyeballs are not perfectly spherical and rotating them would expose different shapes and stretch the skin around them with unique formations. This can provide highly valuable information for inferring gaze points. \gls{name} uses one speaker and four microphones on each side of the glass frame. The speaker emits frequency-modulated continuous-wave (FMCW) acoustic signals with the frequency above 18 kHz towards the eyeballs. The microphones capture the signals reflected by the eyeballs and their surrounding areas, which are used to process and calculate the echo profiles. These echo profiles are fed to a customized deep learning algorithm based on ResNet-18 to predict the gaze point. 

We conducted two rounds of user studies to evaluate the performance of \gls{name}. During the studies, each participant was asked to look at and follow the instruction points on the screen. In the first round of the study, 12 participants evaluated our first hardware prototype, where the microphones and speakers were glued on a glass frame. The average cross-session tracking accuracy was $4.9\degree$. It confirmed the optimal settings of the sensing system, which helped us design the final prototype. The final prototype (as shown in Fig.~\ref{Fig: patterns}) features a more compact form factor, significantly lowers signal strength, and improves environmental sustainability as it can be attached to different glasses. To ensure consistent performance between the two prototypes, we conducted a second round of study with 10 participants, including some new participants, evaluating the final prototype. The final prototype achieved an average tracking accuracy of $4.9\degree$ for cross-session scenarios and $3.6\degree$ for in-session scenarios with a refreshing rate of 83.3 Hz. We made a \href{https://youtu.be/XvNLNkfQY7Q}{\textcolor{blue}{demo video}}\footnote{\url{https://youtu.be/XvNLNkfQY7Q}} to demonstrate the tracking performance and real-world applications of our system.

Although the current accuracy of our system was worse than commercial eye trackers such as Tobii Pro Glasses 3 \cite{tobiiglasses} and Pupil Labs Glasses \cite{pupil}, it is still comparable to some webcam-based eye-tracking systems \cite{webGazer,realEye}. Furthermore, due to the low-power feature of acoustic sensors, \gls{name}, including the data collection system, has a relatively low power signature of 287.9 mW. Compared to camera-based wearable eye tracking systems, our proposed system reduces the power consumption by over 95\%. If using a battery with the capacity similar to Tobii Pro Glasses 3, our system can extend the usage time from 1.75 hours to 38.5 hours. It can even last 6.4 hours on the battery of normal smart glasses, such as Google Glass. The power signature of our system can be further improved using a recently introduced micro-controller with a low-power CNN accelerator (MAX78002). Hence, we implemented our gaze tracking pipeline fully on MAX78002. With the refresh rate set as 30 Hz, the power consumption of the whole system including the data preprocessing and model inference is measured as 95.4 mW.

In summary, the contributions of our paper are as follows: 

\begin{itemize}
    \item We designed and implemented the first acoustic-based continuous eye tracking system on glasses.

    \item A user study with 20 participants showed an average cross-session accuracy of $4.9\degree$ with a refreshing rate up to 83.3 Hz and a power signature of 287.9 mW.

    \item The performance of the system remained robust under different noisy environments and with different styles of glass frames.

    \item A real-time pipeline was implemented on MAX78002 to make inferences on the board with a power consumption of 95.4 mW at 30 Hz.
\end{itemize}

\section{Related Work}
In this section, we introduce the prior webcam-based, non-wearable, and wearable eye tracking systems.

\subsection{Webcam-based Eye Tracking Systems}
Webcams have been widely used to implement eye tracking technologies because of its ubiquity on computers and its advantage of low-cost. Researchers have done many work to explore the potential of webcams for eye tracking \cite{papoutsaki2017searchgazer,papoutsaki2015scalable,saxena2022towards,wisiecka2022comparison}. Currently, there are plenty of webcam-based eye tracking platforms that are available online, such as RealEye.io \cite{realEye}, GazeRecorder \cite{gazeRecorder}, and WebGazer.js \cite{webGazer}.

These webcam-based eye tracking platforms provide affordable solutions for eye tracking with acceptable tracking performance for everyday users. However, the position of webcams are usually fixed and they have a relatively low resolution. Therefore, their performance can be more easily impacted by factors like lighting conditions, occlusions, camera orientations, etc.

\begin{table*}[t]
\caption{\gls{name} and Other Continuous Eye Tracking Techniques. The power of GazeTrak (Teensy 4.1) does not include data preprocessing and deep learning inference. The reported accuracy is tested within the same session without users remounting the device. Both weight and cost include the recording unit. NS = Not Specified.}
\label{Tab: comparision with previous}
\footnotesize
\begin{tabular}{| c | c | c | c | c | c | c | c |}
\hline
\textbf{Reference}  & \textbf{Form Factor} & \textbf{Sensors} & \textbf{Power} & \textbf{Accuracy} & \textbf{Refresh Rate} & \textbf{Weight} & \textbf{Cost}\\
\hline\hline
Cho et al. \cite{cho2012gaze} & Glasses & Cameras & \textgreater7W & $0.79\degree$ & NS & NS & NS\\
\hline
Ryan et al. \cite{ryan2008limbus} & Glasses & Cameras & \textgreater1.6W & $2\degree$ & NS & NS & \textasciitilde\$700\\
\hline
iShadow \cite{mayberry2014ishadow} & Glasses & Cameras & 0.07W & $3\degree$ & 30 Hz & NS & NS\\
\hline
CIDER \cite{mayberry2015cider} & Glasses & Cameras & 0.032W & $0.6\degree$ & 250 Hz & NS & NS\\
\hline
Pupil Labs Glasses \cite{pupil} & Glasses & Cameras & 8.6W & $0.6\degree$ & 30/60/120 Hz & 202.75g & \$2,849\\
\hline
Tobii Pro Glasses 3 \cite{tobiiglasses} & Glasses & Cameras & 10.7W & $0.6\degree$ & 50/100 Hz & 388.5g & \$16,055 \\
\hline
SMI Glasses \cite{smi} & Glasses & Cameras & NS & $0.5\degree$ & 60/120 Hz & NS & \$41,000\\
\hline
Li et al. \cite{li2018battery} & Glasses & NIR LED \& Photodiodes & 395\micro W & <2\degree & 120 Hz & <25g & NS\\
\hline
Li et al. \cite{li2017ultra} & Head-mounted & Photodiodes & 791\micro W & 6.3\degree & 10 Hz & NS & NS\\
\hline
GazeRecorder \cite{gazeRecorder} & Webcam & Camera & / & $1.05\degree$ & 30 Hz & / & \$500/month\\
\hline
WebGazer.js \cite{webGazer} & Webcam & Camera & / & $4.17\degree$ & NS & / & Free\\
\hline
RealEye.io \cite{realEye} & Webcam & Camera & / & \textasciitilde$5\degree$ & 60 Hz & / & \$600/month\\
\hline
\textbf{\gls{name} (Teensy 4.1)} & \textbf{Glasses} & \textbf{Acoustic Sensors} & \textbf{0.288W} & $\textbf{3.6\degree}$ & \textbf{83.3 Hz} & \textbf{44.2g} & \textbf{\textasciitilde\$75}\\
\hline
\textbf{\gls{name} (MAX78002)} & \textbf{Glasses} & \textbf{Acoustic Sensors} & \textbf{0.095W} & $\textbf{4.2\degree}$ & \textbf{30 Hz} & / & /\\
\hline
\end{tabular}
\end{table*}

\subsection{Other Non-wearable Eye Tracking Technologies}

In order to provide more accurate and reliable solutions to eye tracking and make them more applicable to assorted scenarios, researchers have put lots of efforts in implementing other non-wearable eye tracking technologies other than webcam-based systems, most of which are based on cameras with a higher resolution than webcams. Frontal camera-based eye tracking technologies based on computer vision techniques can take full advantage of the whole facial information of the user for eye tracking, which usually leads to high tracking accuracy. Different kinds of cameras have been used in tracking eye movements, such as RGB cameras \cite{ablavatski2020realtime}, infrared (IR) cameras \cite{ohno2003just}, and thermal cameras \cite{wang2016thermographic}. Beyond using just one camera, many technologies adopted multiple cameras in their eye tracking systems in order to improve the performance in different perspectives including providing larger tracking coverage \cite{ahlstrom2010comparison}, allowing for user motion \cite{hennessey2012long} and tracking eye movements of multiple users \cite{mahanama2022multiuser}. Because of the reliable tracking performance and reasonable calibration time needed, frontal camera-based eye tracking technologies have been well commercialized, among which Tobii Pro Fusion \cite{tobii} is one of the best desktop eye trackers because it only requires seconds of calibration process for new users and can provide a tracking accuracy as low as $0.3\degree$ in optimal conditions. As a result, this product has been used as reference in many research projects.

The aforementioned frontal camera-based technologies are mostly located at fixed positions and do not work well while users move to another position or are walking around. In order to allow some mobility for users while they are using the eye tracking technologies, many researchers investigated utilizing the cameras on mobile devices to track eye movements, such as mobile phones \cite{zhu2018mobiet,lei2021eye,krafka2016eye} or tablets \cite{bafna2021eyetell,krafka2016eye}. However, these eye tracking technologies based on mobile devices still require users to hold the mobile devices in front of their face all the time and cannot provide completely hands-free and motion-free experiences for users.

\subsection{Wearable Eye Tracking Technologies}
To overcome the challenges that non-wearable eye tracking technologies face as described in the last two subsections, many wearable eye tracking technologies based on cameras \cite{zhang2011discrimination, ryan2008limbus, noris2011wearable,cho2012gaze,kassner2014pupil,lanata2015robust,mayberry2014ishadow,mayberry2015cider}, optical sensors \cite{borsato2016episcleral,topal2008head,topal2008wearable,li2018battery,li2017ultra}, acoustic sensors \cite{golard2021ultrasound}, magnetic sensors \cite{whitmire2016eyecontact}, Electrooculography (EOG) sensors \cite{bulling2008s, bulling2009wearable}, or inertial measurement units (IMU) \cite{lanata2015robust} have been deployed on different kinds of wearables including glasses \cite{topal2008head,topal2008wearable,zhang2011discrimination, borsato2016episcleral, ryan2008limbus,cho2012gaze,golard2021ultrasound,li2018battery,mayberry2014ishadow,mayberry2015cider}, goggles \cite{bulling2008s, bulling2009wearable}, hat \cite{lanata2015robust}, and head-mounted devices \cite{whitmire2016eyecontact, noris2011wearable, kassner2014pupil,li2017ultra}. Among all these wearable eye tracking technologies, camera-based ones usually outperform others in terms of tracking performance and do not require lots of calibration data from new users. Many wearable eye trackers using cameras, especially on glasses, have become commercial and can be used as a reliable way to track eye movements continuously, such as Tobii Pro Glasses 3 \cite{tobiiglasses}, Pupil Labs (Invisible, Core, VR/AR add-ons) \cite{pupil}, Dikablis Glasses 3 \cite{dikablis}, and SMI Eye Tracking Glasses \cite{smi}. With these technologies, various novel gaze-based applications have been enabled, including detection of eye contacts \cite{ye2012detecting}, interaction with devices \cite{kyto2018pinpointing, matsubara2015extraction,paletta2014smartphone}, and monitoring mental health \cite{vidal2012wearable, li2020identification}.

Despite of the promising tracking performance, current solutions to wearable eye tracking systems still have some limitations. First of all, many eye tracking systems above can only recognize discrete gestures \cite{zhang2011discrimination,topal2008wearable,topal2008head,bulling2008s,bulling2009wearable}, limiting their performance in applications that need continuous tracking of the eyes. Camera-based wearable eye trackers can provide high accuracy in continuous eye tracking, but cameras are usually power-hungry, which makes them relatively impractical while deployed in wearables that need to be worn in everyday settings. To address this issue, Mayberry et al.~\cite{mayberry2014ishadow,mayberry2015cider} proposed low-power solutions to tracking gaze positions with cameras on glasses while maintaining promising accuracies. Despite of the impressive performance, changing lighting conditions can still be a problem for these camera-based systems, as the performance became worse in an outdoor setting~\cite{mayberry2015cider}. Besides, commercial eye trackers are usually expensive and do not provide open-source software for users, preventing them from being easily accessed and adapted by general users. 

Recently, Li et al. \cite{li2018battery} proposed a low-cost and battery-free solution to continuous eye tracking using near infrared emitters and receivers on glasses. It achieves competitive performance but they stated in the paper that this system can be impacted by direct sunlight and glasses movement, i.e. the remounting of the glasses. Besides, this work tracks the position and size of the pupil so we cannot directly compare it to our system. After conversion, its tracking accuracy of gaze positions is smaller than $2\degree$ in angular error. Another system using similar technology from the same group \cite{li2017ultra} tracks gaze positions with an accuracy of $6.3\degree$, worse than the tracking accuracy of our system at $4.9\degree$. Golard et al. \cite{golard2021ultrasound} conducted a modeling and empirical study to prove that ultrasound can provide a low-power, fast and light-insensitive alternative for camera-based eye tracking technologies. However, it was evaluated on a physical 3D model of a human eye and used time-of-flight estimated from acoustic signals and not clear how it can apply on a real user. 

To the best of our knowledge, \gls{name} is the first wearable sensing technology based on active acoustic sensing that can track gaze points continuously. We summarized and compared \gls{name} with some aforementioned wearable and webcam-based eye tracking techniques that can continuously track gaze positions in Tab.~\ref{Tab: comparision with previous}. These techniques are those that are most related to our system. Please note that commercial eye tracking wearables \cite{tobiiglasses,pupil,smi} usually have camera(s) recording the video of the environment as well so we can only roughly compare them to our device in terms of power and weight.

\begin{figure*}[t]
  \includegraphics[width= 0.9 \textwidth]{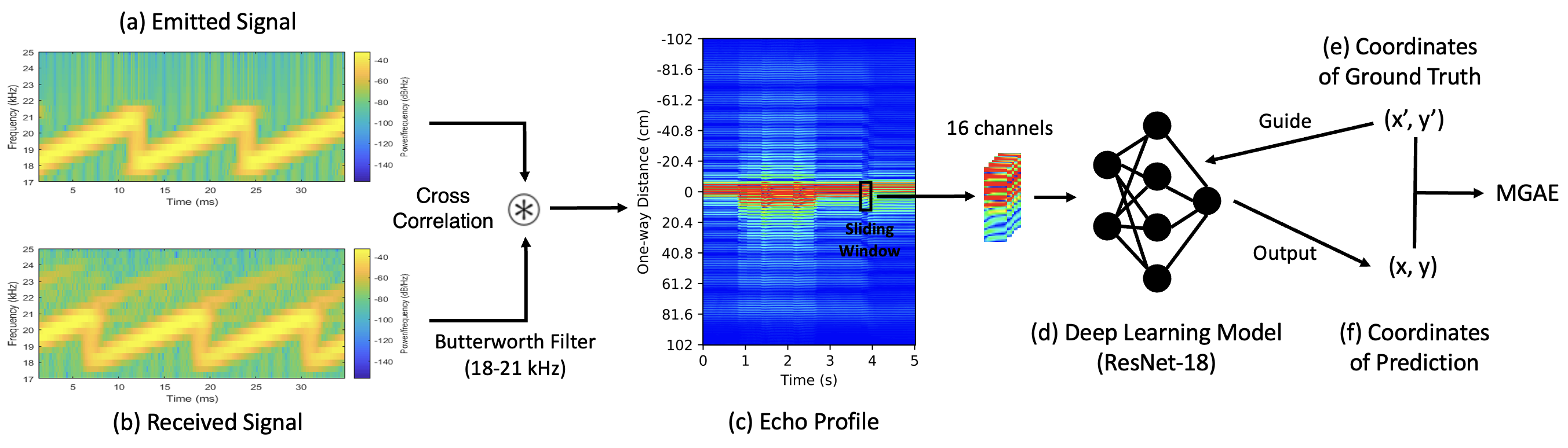}
  \caption{Overview of the \gls{name} System: Use the Speaker on the Right Side (18-21 kHz) for Illustration.}
  \label{Fig: flowchart}
\end{figure*}

\section{Principle and Algorithms}
\label{Sec: Principle}
Active acoustic sensing is based on affordable sensors (speakers and microphones), the sizes of which are relatively small. Previous research work has proved that it is able to provide enough information to track subtle skin deformations such as facial expressions \cite{li2022eario,Gao2022sonicface}. In this section, we discuss how this approach can be adapted to eye tracking.

\subsection{FMCW-based Active Acoustic Sensing}
\label{SubSec: FMCW}
In order to capture the formation around eyeballs, we use FMCW-based acoustic sensing, which has been widely proven effective to estimate distance and movements from complex environments \cite{nandakumar2015contactless,wang2018c}. 

\subsubsection{Encoded FMCW Signals}
\label{Subsubsec: FMCW signals}

While customizing the FMCW signals for our system, three main features are taken into account: 1) \textit{Operating frequency range}: The device is expected to be worn by users for a long period of time in their everyday lives. As a result, the FMCW signals need to be transmitted in the inaudible frequency range. Besides, to ensure the encoded signals are minimally impacted by the noise in the environment, the operating frequency range we pick should also be uncommon in daily settings; 2) \textit{Sampling rate}: To achieve a reasonable spatial and temporal resolution of tracking eye movements, the sampling rate of FMCW signals must be high enough; 3) \textit{Gain}: As power signature increases with the signal gain, the signal gain should be properly determined to balance signal strength and power consumption.

Considering all the factors above, we set the operating frequency range of the FMCW signals that we emit in the \gls{name} system above 18 kHz, because this range is near-inaudible and uncommon in the sounds generated by normal human activities. Because both eyes contain information while moving, we placed one speaker on each side of the glass frame. We set the speaker on the right side to operate at 18-21 kHz while the one on the left side operates at 21.5-24.5 kHz to make sure they do not interfere with each other. To guarantee that the system works reliably in these frequency ranges, we set the ADC sampling rate as 50 kHz with the frame length of FMCW signals as 600 samples. This gives the system a refresh rate of eye tracking at 83.3 Hz (50000 samples/s $\div$ 600 samples). 
We believe a refresh rate of 83.3 Hz is sufficient to provide continuous gaze tracking since the frame rate of most videos are 30 Hz or 60 Hz. Lastly, the gain was experimentally adjusted to make sure that the signal does not saturate the microphones while the power consumption is relatively low. 

\subsubsection{Acoustic Patterns for Continuous Eye Tracking}

\label{SubsubSec: Echo Profiles}
After receiving the reflected FMCW signals, we first apply a Butterworth band-pass filter with a cut-off frequency range of 18-21 kHz or 21.5-24.5 kHz on the signal to remove the signals in the frequency range that we are not interested in. It also helps protect the privacy of users because we remove the audible range of the signals. Then we further process the filtered signal to obtain unique acoustic patterns. According to prior research work \cite{wang2018c,li2022eario,zhang2023echospeech,sun2023echonose,mahmud2023posesonic,zhang2023hpspeech,li2024eyeecho,lee2024echowrist}, \textit{Echo Profile} provides an accurate depiction of the status and movements of the reflecting objects in the environment. As a result, in this paper, we also use echo profiles as the acoustic patterns that our system monitors. As shown in Fig.~\ref{Fig: flowchart} (a)-(c), echo profile is obtained by continuously calculating the cross-correlation between the received signals and transmitted signals. Fig. \ref{Fig: patterns} demonstrates that different eye fixations and movements are correlated with different patterns in echo profiles. Based on these observations above, we believe that our \gls{name} system utilizing FMCW-based active acoustic sensing is able to track eye movements continuously with high accuracy.

\subsection{Machine Learning Algorithms}
\label{SubSec: Machine Learning}

\subsubsection{Ground Truth Acquisition without using Eye Trackers} 
\label{Subsubsec: ground truth}
A professional eye-tracker (e.g., Tobii Pro Fusion) can provide highly accurate ground truth, but it is expensive. If our system needs a professional eye-tracker to train the system, it will make our eye-tracking system less accessible. 

Therefore, we developed a new ground truth acquisition and calibration system that only needs a program running on a laptop. The program generates instruction points on the screen as the ground truth. During data collection, the users only need to look at and follow the movements of the instruction points. These ground truth data along with the echo profiles are fed into the machine learning model for training. This method is generally applicable on any device with a screen. For details about how the instruction points are generated, please refer to Sec.~\ref{Sec: user study}. To better compare our system with commercial eye trackers, we also use a Tobii Pro Fusion (120 Hz) \cite{tobii} to record the eye movements to demonstrate the effectiveness of our training methods. 

\subsubsection{Deep Learning Model}
We developed a customized deep-learning pipeline to learn the echo profiles calculated on the received signals. Because in the echo profiles (See Fig. \ref{Fig: patterns}), the temporal information has been converted to the spatial information on an image, we decided to use ResNet-18 as the encoder of our deep learning model because CNN networks are known to be good at extracting features from images. Then a fully-connected network is used as a decoder to predict gaze positions based on the features extracted from the images. 

\begin{figure*}[t]
  \includegraphics[width= 1 \textwidth]{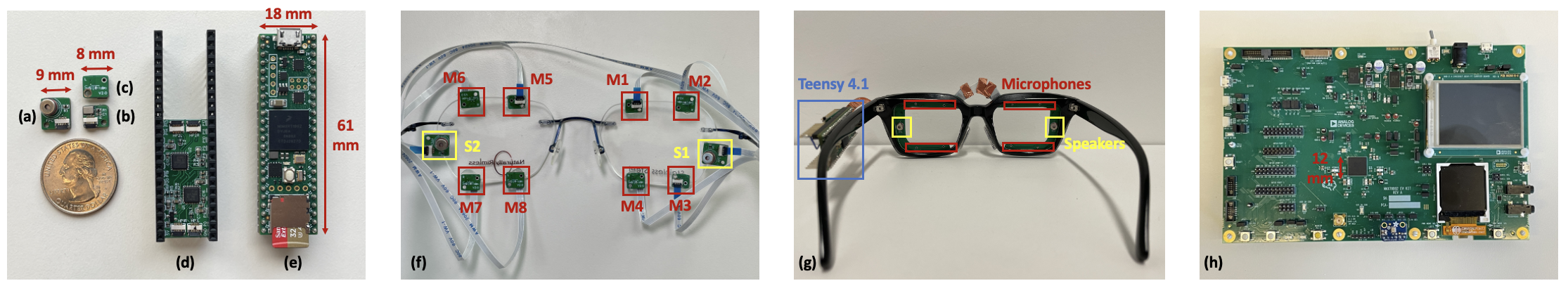}
  \caption{Hardware and Form Factor for GazeTrak: (a) Speaker board; (b) Microphone board (front view); (c) Microphone board (back view); (d) Customized PCB board for the audio chip NXP SGTL5000; (e) Teensy 4.1; (f) Glasses form factor with speakers and microphones attached (M1-8: microphones, S1-2: speakers); (g) Attachable and more compact prototype; (h) MAX78002 Evaluation Kit.}
  \label{Fig: hardware}
\end{figure*}

Because of the limited distance between the sensors on the glasses and the eyes, we are only interested in a certain range of the echo profiles (Fig.~\ref{Fig: flowchart} (c)). As a result, we crop the echo profiles of each channel to get the center 70 pixels (23.8 cm) vertically. Then we randomly select 60 consecutive pixels (20.4 cm) out of these 70 pixels for data augmentation purpose to make sure the system will not be severely impacted by the vertical shifting caused by remounting the device. To continuously track the gaze positions, we apply a sliding window of 0.3 seconds on the echo profiles. As a result, the dimension of the echo profile that we input into the deep learning model for one channel is 26 (0.3 s $\times$ 50000 Hz $\div$ 600 samples + 1) $\times$ 60 (pixels). Because we use 2 speakers and 8 microphones in our system, which will be illustrated in Subsec. \ref{Subsec: form factor}, we crop out the same dimension of echo profiles for all 2 $\times$ 8 = 16 channels, making the dimension of the input vector to the deep learning model as 26 $\times$ 60 $\times$ 16. 

We use the instruction points as the labels (see Subsubsec. \ref{Subsubsec: ground truth}) and the mean squared error (MSE) as the loss function. We chose Adam optimizer and set the learning rate as 0.01. The model is trained for 30 epochs to get the estimation of the two gaze coordinates (x, y).

\subsubsection{Evaluation Metrics}
\label{Subsubsec: evaluation metrics}
The prediction of our system is the coordinate (x, y) of our estimated gaze position on the screen in pixels. To evaluate the accuracy of \gls{name}, we adopted the accuracy defined in COGAIN eye tracker accuracy terms and definitions \cite{cogain}. The evaluation metric we use in our system is the mean gaze angular error (MGAE) between the coordinate of our prediction (x, y) and that of the ground truth (x', y'). To calculate MGAE in degrees from the coordinates, we first need to get the angular error $\theta$ between the prediction and the ground truth of each data point. $\theta$ can be calculated using the law of cosines in a triangle as follows:

\begin{equation}
\label{Eq: angle}
    \theta = \arccos{(\frac{d_{eg}^2+d_{ep}^2-d_{gp}^2}{2 \times d_{eg} \times d_{ep}})} \times 180 \div \pi
\end{equation}
where $d_{eg}$, $d_{ep}$ and $d_{gp}$ are the distance between user's eyes and ground truth, the distance between user's eyes and prediction, and the distance between ground truth and prediction respectively. MGAE is obtained by averaging $\theta$ over all the data points in the testing dataset.
\section{Design and Implementation}
\label{Sec: Implementation}

\subsection{Hardware Design}

In order to implement the FMCW-based active acoustic sensing technique mentioned in the section above, we chose Teensy 4.1 \cite{teensy} as the micro-controller to provide reliable FMCW signal generation and receiving in multiple channels. We designed a PCB board to support two SGTL5000 chips which are the same as the one on the Teensy audio shield \cite{teensyaudio}. With this customized PCB board plugged onto Teensy, it can support as many as 8 microphones and 2 speakers. We chose the speaker called OWR-05049T-38D \cite{owr} and the MEMS microphone called ICS-43434 \cite{ics} to support signal transmission and reception. We also built customized PCB boards for the speaker and the microphone to make them as small as possible. We used the Inter-IC Sound (I2S) buses on the Teensy 4.1 to transmit data between the Teensy 4.1 and the SGTL5000 chips, speakers and microphones. The collected data is stored in the SD card on Teensy 4.1. Fig.~\ref{Fig: hardware} (a)-(e) show these components. 

\subsection{Form Factor Design}
\label{Subsec: form factor}

We designed the first form factor using a commodity glass frame. We glued 1 speaker and 4 microphones to each inner side of a pair of light-weight glasses. The speakers and microphones are symmetrically placed on the glasses, as shown in Fig. \ref{Fig: hardware} (f). 

Based on the experience we learned during the iteration process, there are three key factors we took into consideration while designing the final form factor of GazeTrak: 1) \textit{Type of glass frame}: We started designing the form factor with a large glass frame because we believe it has more room for us to place sensors. However, the larger the glass frame is, the easier it will be for the frame to touch the skin, blocking the signal transmission and reception. As a result, we finally picked a relatively small glass frame with a nose pad that can support the glass frame to a higher position. Besides, the light-weight glasses minimize the pressure attached on the user's nose, making it more comfortable to wear; 2) \textit{Sensor position}: The speakers and microphones on two sides are symmetric because we believe the movements of two eyes are usually synchronized. On each side, we place the speaker on the frame of the glasses next to the outer canthi because it is easier for the speakers to touch the skin if they are placed above the cheekbones or next to the eyebrows, considering their height. The microphones are scattered on the frame as far away from each other as possible to capture more information by receiving signals travelling in different paths. The sensors are attached as far away from the center of the lenses as possible in order to avoid blocking the view of the user; 3) \textit{Stability}: We found that the stability of the device severely impacts the performance of our system especially when users need to remount the device frequently. The anti-slippery nose pad prevents the glasses from sliding down the user's nose. Furthermore, we added two ear loops at the end of the legs of the glass frame. They greatly helps fix the glasses position from behind ears and improves the performance of the system. Finally, we made the form factor as shown in Fig.~\ref{Fig: hardware} (f). 

\subsection{Final Hardware Prototype}
\label{Subsec: attachable prototype}
The prototype above is suitable for initial testing and comparison of different configurations. However, once the design of the prototype is finalized, we aim to create a more compact and less obtrusive form factor that is suitable for everyday use by users. To achieve this, we have designed two PCB boards, each containing one speaker and four microphones onboard, which can be attached to one side of the glasses. We have also deployed the Teensy 4.1 and the PCB board with SGTL5000 chips directly onto one leg of the glasses. To connect the micro-controller and the customized PCB boards, we have used flexible printed circuit (FPC) cables. The system has an interface that allows it to be powered by a Li-Po battery. The compact prototype is shown in Fig.~\ref{Fig: hardware} (g), and Fig.~\ref{Fig: patterns} shows a user wearing the prototype. We believe that this prototype can be easily adapted and attached to different types of glasses.

We have measured the weight of the prototype, and it carries a total weight of 44.2 grams, including the glasses, Teensy 4.1, PCB boards, and the Li-Po battery. Compared to camera-based eye tracking glasses, our \gls{name} device is much lighter. For example, Tobii Pro Glasses 3~\cite{tobiiglasses} weigh 76.5 grams for the glasses and 312 grams for the recording unit. Our device has a significant advantage over camera-based eye tracking glasses in terms of weight.

\section{User Study Procedure}
\label{Sec: user study}

The objective of our user study is to validate the performance of \gls{name} on continuously tracking gaze points. In order to reach this goal, we carefully designed the instruction video for participants' gaze to follow. Basically, on the white screen, there would be one red dot moving around and we asked participants to stare at the point and follow it with their eyes. We divided the screen into 100 regions. For each data point, the instruction point appeared at a random position within one random region. The instruction point would move quickly to that random position and stay static at that position for a certain period of time because we mainly would like to test how \gls{name} performs to track the fixation of participants. 

We recruited 20 participants (10 females and 10 males, 22 years old on average). Note that some participants participated in the study for multiple times to test different settings. The study was conducted in an experiment room on a university campus. During the study, the participants sit on a chair and put on the glasses form factor with our \gls{name} system. For each participant, we produced 12 sessions of instruction points. During the interval between sessions, participants were instructed to remove the device, place it on the table, and then put it back on. This step was taken to demonstrate that our system continued to function correctly even after the device was remounted. In each session, the instruction point moved to all the 100 pre-defined regions in a random order. The duration for which the instruction point remained at each position varied from 0.5 to 3.5 seconds, with an average of 2 seconds. As a result, the average length of each instruction session was 200 seconds. Before each session, there was a 15-second calibration process with the instruction point moving to the four corners of the screen and the center of the screen. 

The full study took no more than 1.5 hours for each participant, during which we collected approximately 40 minutes of data (200 seconds $\times$ 12 sessions). Upon completing the study tasks, the participant was asked to complete a questionnaire to collect their demographic information and their feedback using this system.
\section{Evaluation Results}
\label{Sec: Results}
In this section, we first evaluated the performance of \gls{name} with the initial prototype, comparing different ground truth acquisition methods, sensor configurations and amounts of training data. Then we tested our system under noisy environments and on glasses of various frame styles. Finally, we optimized the system on the final prototype and evaluated it with another study, with power consumption measured.

\begin{table*}[h]
\small
\caption{Study Results for Different Mic Configurations.}
\label{Tab: different mics}
\scalebox{0.8}{\begin{tabular}{| c | c | c | c | c | c | c | c | c |} 
\hline
 Mic Configuration & M1+M5 & M2+M6 & M3+M7 & M4+M8 & Best 4 Mics (M2,4,6,8) & Best 6 Mics (M1,2,4,5,6,8) & All 8 Mics\\ [0.5ex] 
 \hline\hline
 MGAE & $7.7\degree$ & $7.2\degree$ & $8.5\degree$ & $6.9\degree$ & $5.9\degree$ & $5.5\degree$ & $4.9\degree$\\
 \hline
\end{tabular}}
\end{table*}

\subsection{User-dependent Model}
\label{Subsec: user-dependent}
We first tested our system using the first prototype in Fig.~\ref{Fig: hardware} (f) with 12 participants. A user-dependent model was applied to train a separate model for each participant. Among 12 sessions we collected for each participant, we conducted a 6-fold cross validation to test the tracking performance of our system by using 10 sessions (33.3 minutes) of data for training and 2 sessions (6.7 minutes) of data for testing. Using the evaluation metrics defined in Subsubsec.~\ref{Subsubsec: evaluation metrics}, we calculated the mean gaze angular error (MGAE) in degree for all participants, and the result we obtained was $5.9\degree$. To further improve the performance, we adopted the 15-second calibration data before each session to fine-tune the model, which resulted in an improved performance of $4.9\degree$. It is worth mentioning that a similar calibration process is also required for commodity eye trackers (e.g., Tobii Pro). The distance between the participants' eyes and the screen center is measured to be around 60 cm so we have a field of view of $60\degree$ (the largest possible angular error) in this study. We made a \href{https://youtu.be/XvNLNkfQY7Q}{\textcolor{blue}{demo video}} showing how our prediction looks like visually with this level of accuracy.

\begin{figure}[h]
 \centering
  \includegraphics[width=0.4\textwidth]{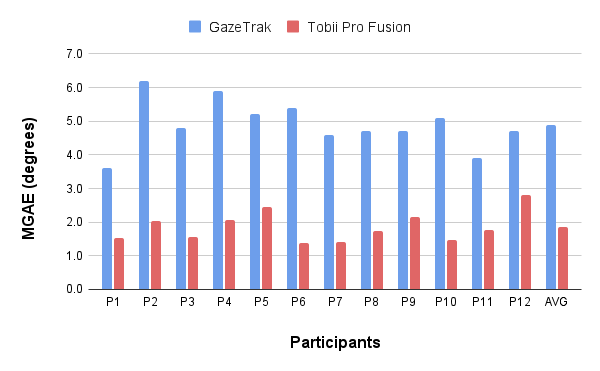}
  \caption{MGAE Distribution across Participants.}
  \label{Fig:MGAE_Participants}
\end{figure}

Next, we aimed to compare the impact on gaze tracking performance of using different ground truth acquisition methods: a commodity eye tracker (Tobii Pro Fusion) versus our method (using instruction points on the screen). We used the eye tracking data recorded by Tobii Pro Fusion as the ground truth to train the model, and the MGAE after fine-tuning was $4.9\degree$. We conducted a repeated measures \textit{t}-test between the results using Tobii data as the ground truth and those using instruction points as the ground truth for all 12 participants, and did not find a statistically significant difference ($p = 0.92 > 0.05$). This suggests that using instruction points on a screen monitor as the ground truth can be as effective as using Tobii data.

Apart from that, we also recorded the eye tracking accuracy of Tobii Pro Fusion itself which was reported after the calibration process of the Tobii platform. The results showed that Tobii Pro Fusion can track the gaze points with an average accuracy of $1.9\degree$ during the calibration process for all participants. We plotted the tracking performance of both \gls{name} and Tobii for all participants in Fig.~\ref{Fig:MGAE_Participants}.

\subsection{Impact of Sensor Configurations}
In this subsection, we evaluated the impact of the number and placement of microphones on tracking performance to determine the optimal sensor position for the best results. We assessed four different settings: 1) one microphone on each side (left and right); 2) two microphones on each side; 3) three microphones on each side and 4) all four microphones on each side. In the first setting, we compared the performance using data from four sets of microphone settings (M1+M5, M2+M6, M3+M7, and M4+M8 in Fig.~\ref{Fig: hardware} (f)), which is presented in Tab. \ref{Tab: different mics}. The findings demonstrate that the M4+M8 pair of microphones provides the best tracking performance among the four pairs tested. We conducted a one-way repeated measures ANOVA test on the results of the four settings and identified a statistically significant difference ($F(3, 44) = 6.74, p = 0.001 < 0.05$). These results indicate that microphone placement can affect gaze tracking performance, possibly due to differences in signal reflection before arriving at different microphones.

We further conducted experiments to evaluate performance using different combinations of microphones under settings 2 and 3. The results showed that the best performance was 5.9 degrees and 5.5 degrees, respectively. We also ran a one-way repeated measures ANOVA test among the results of these four settings using data from 12 participants. The results showed a statistically significant difference ($F(3, 44) = 51.61, p = 0.00001 < 0.05$). These findings suggest that our system requires four microphones on each side (eight microphones in total) to achieve the best performance.

\subsection{Impact of Blinking}
\label{Subsec: blink impact}

Blinking can introduce noise in our highly-sensitive acoustic sensing system as it can lead to relatively large movements around the eye. We conducted an evaluation to determine whether blinking affects the tracking performance of our system. For this evaluation, we selected data from three participants with the best, worst, and average tracking performance (P1, P2, P10). We removed the data where the participant blinks (about 10\% of total data) based on the ground truth data obtained from Tobii Eye Tracker. We then used the processed data to retrain the user-dependent model for each participant. Our results showed that the performance did not improve after removing the blinking data. One possible reason for this result is that the blinking patterns are consistent and can be learned by the machine learning model. Therefore, our findings suggest that blinking does not significantly impact the performance of our system.

\subsection{User-adaptive Model}
\label{Subsec: user-adaptive}
To reduce the need of providing lots of training data for a new user, we employed a three-step process to train a user-adaptive model. Firstly, we trained a large base model using data from all participants except the one being tested. Secondly, we fine-tuned the model using the training data collected from the current participant. Notably, the user only needs to provide training data once during the initial system use. Finally, at the beginning of each session, we further fine-tuned the model using calibration data collected from the participant before testing or using the system. To determine the amount of data required to achieve competitive tracking performance, we reserved two sessions of data for testing and used varying amounts of training data from the participant to fine-tune the large model.

The results show that a new user only needs to provide six sessions of training data (approximately 20 minutes) to achieve good performance. Collecting more data does not necessarily result in better performance. Additionally, with only two or three sessions of data (approximately 6 minutes), the system can achieve a performance of $6.7\degree$ and $6.1\degree$, respectively. If no user data is collected, the performance is $11.3\degree$. This is likely because different people have unique head, face, and eye shapes. Therefore, to further reduce the amount of training data required from each new user, we may need to collect a significantly larger amount of training data from a more diverse set of participants.

\subsection{Impact of Environment Noise}

To ensure that our acoustic sensing system is resistant to different types of environmental noise, we conducted two experiments as described in this subsection. 

\subsubsection{Noise Injection}
\label{Subsubsec: noise injection}
In the first experiment, we recorded noises in different environments using the microphones on our glass frame. We then overlaid the noise onto the data collected in the user study to simulate different noisy environments.
We recorded the noise in four different environments and
measured the average noise levels using a sound level meter app called NIOSH provided by CDC \cite{niosh}: 1) \textit{street noise (70.8 dB(A))} recorded on the street near a crossroad; 2) \textit{music noise (64.5 dB(A))} recorded while playing music on a computer; 3) \textit{cafe noise (54.5 dB(A))} recorded in a cafe; 4) \textit{driving noise (65.6 dB(A))} recorded while driving a vehicle. After overlaying each of these four noises, the tracking performance remained unchanged for every participant.

\subsubsection{Real-world Noisy Environments}
\label{Subsubsec: study noisy}
In the second experiment, we invited eight participants from the previous user study and recruited two new participants (P13 and P14) to test our device in different real-world noisy environments. Since this study required us to move to different environments, the study design differed slightly from the previous study described in Sec. \ref{Sec: user study}.

In this study, we used an Apple MacBook Pro with a 13.3-inch display to play the instruction videos. We used the instruction points as the ground truth. The MacBook Pro was placed on a movable table, and participants were instructed to sit in front of the table to conduct the study. Additionally, according to Subsec.~\ref{Subsec: user-adaptive}, 6 sessions of training data are sufficient to provide acceptable tracking performance. Therefore, for each participant, we collected a total of 8 sessions of data in a quiet experiment room, with 6 sessions for training and 2 sessions for testing. We then collected additional testing data under two different noisy environments. In the first environment, participants used our system while we played random music for 2 sessions. In the second environment, we collected 2 sessions of testing data at a campus cafe where staff and people were talking around during business hours. The noise levels under each environment were measured using the CDC NIOSH app: 1) \textit{quiet room (33.8 dB(A))}; 2) \textit{play music (64.0 dB(A))}; 3) \textit{in the cafe (56.6 dB(A))}. This study design led to a total of 12 sessions of data collection for each participant, which is the same as the previous study.

We trained a personalized model for each participant using 6 sessions of data collected in the quiet room. Then, the 2 testing sessions collected in each scenario were used to test the performance of our system in different environments. The average gaze tracking performance of our system across 10 participants remained satisfactory at $3.8\degree$ and $4.8\degree$ under two noisy environments, playing music and in the cafe, while the performance in the quiet room was $4.6\degree$. Overall, the average accuracy of gaze tracking did not change significantly with the presence of noise in the environment. We conducted a one-way repeated measures ANOVA test among the results of these three scenarios for all 10 participants and did not find a statistically significant difference ($F(2, 27) = 2.46, p = 0.11 > 0.05$). This again validates that our system is not easily affected by environmental noise.

\begin{figure*}[h]
    \centering
    \subfloat[Original Glasses (F1)]{
        \includegraphics[height=.15\textwidth]{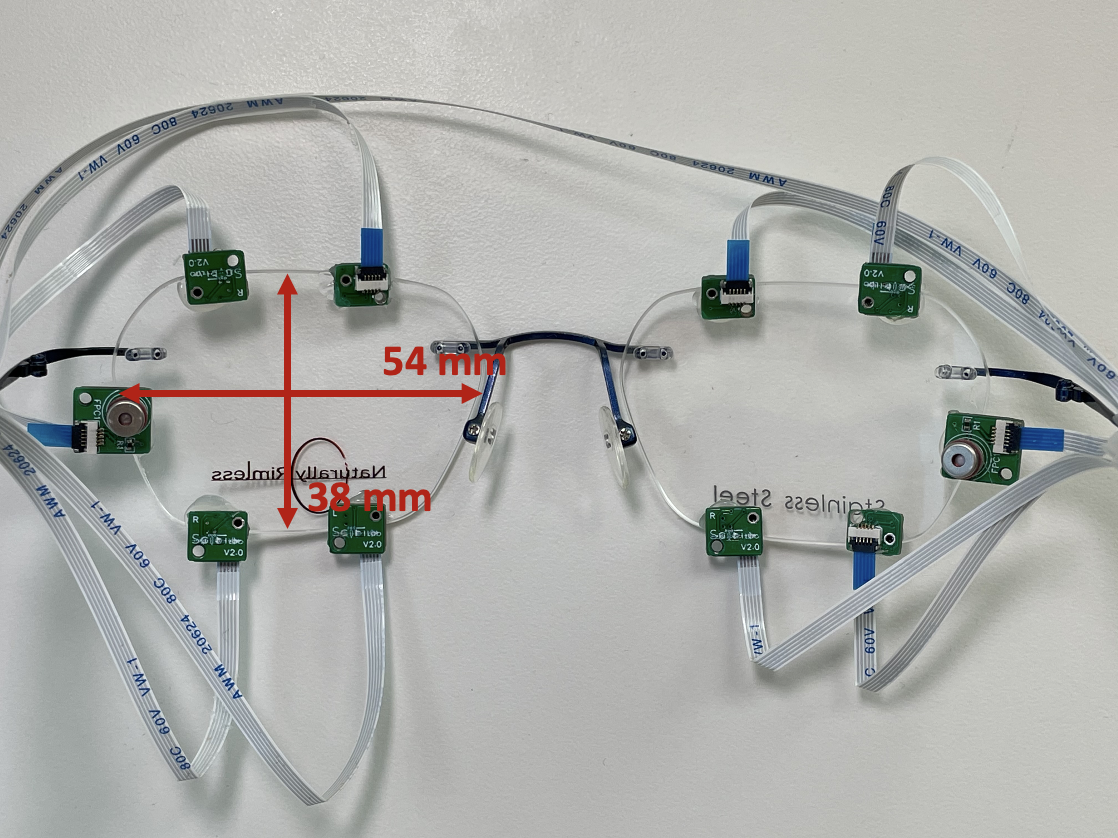}
    }
    \hspace{.08\textwidth}
    \subfloat[Small Glasses (F2)]{
        \includegraphics[height=.15\textwidth]{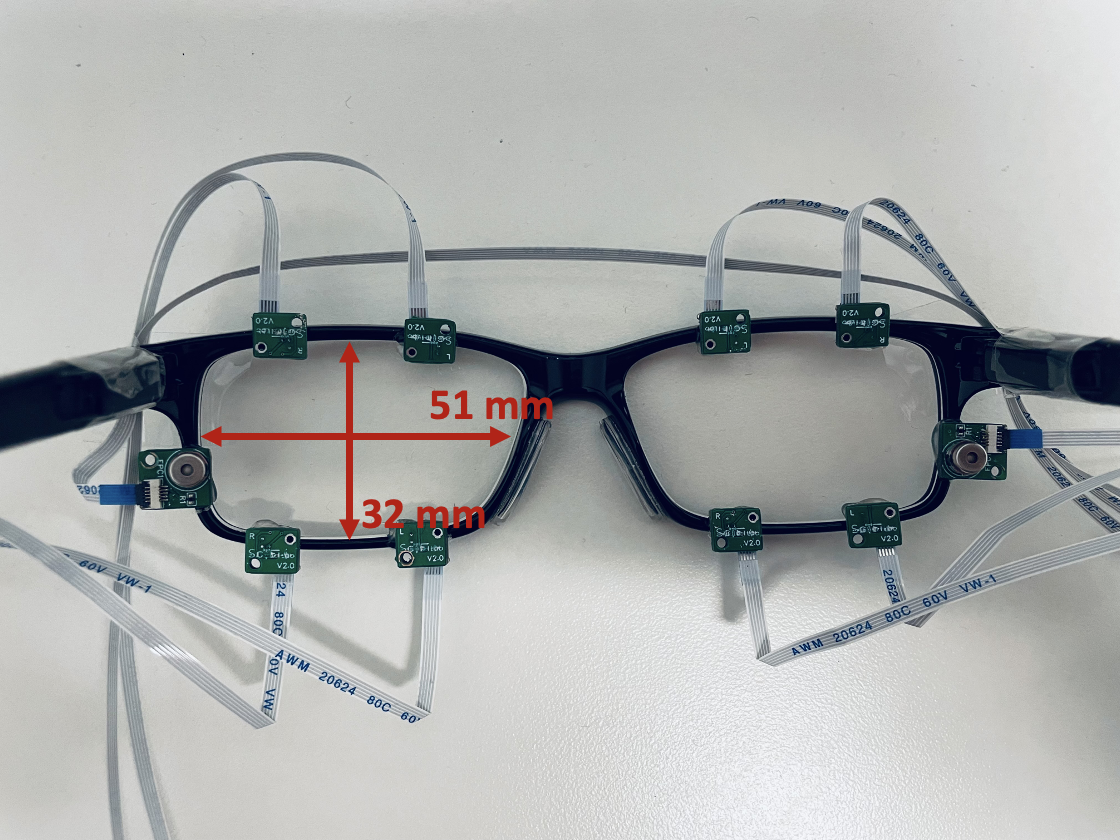}
    }
    \hspace{.08\textwidth}
    \subfloat[Large Glasses (F3)]{
        \includegraphics[height=.15\textwidth]{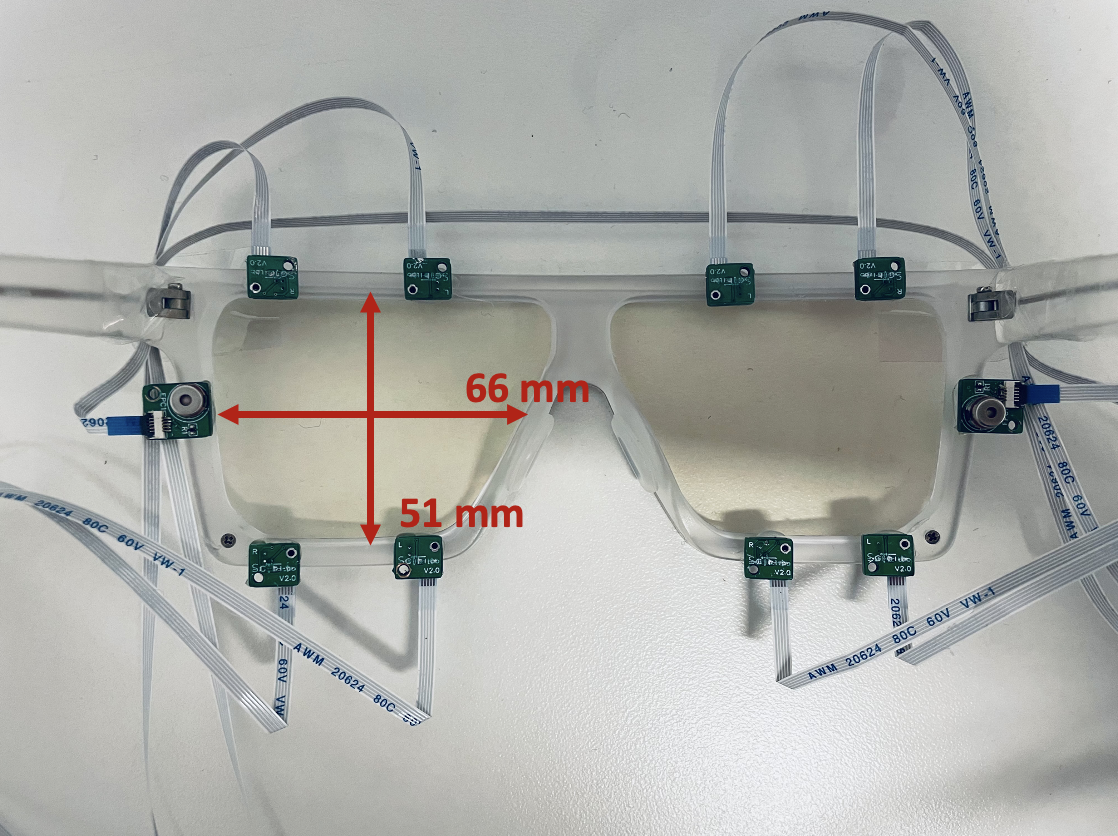}
    }
    \caption{\gls{name} Deployed on Glasses with Various Frame Styles.}
    \label{Fig: different glasses}
\end{figure*}

\subsection{Impact of Different Glass Frames}
In our user study, we only tested our system on one glass frame (F1). However, we believe our \gls{name} system can be easily applied to glasses with different frame styles. In order to validate this assumption, we deployed our system on two other pairs of glasses as shown in Fig.~\ref{Fig: different glasses}. The original glass frame in Fig.~\ref{Fig: different glasses} (a) is frameless and relatively lightweight. In this study, we applied our system on two new glasses with different styles, size and weight. The first new glass frame (small glasses, F2) has a smaller size than F1 but a larger weight due to the frame around the lens (Fig.~\ref{Fig: different glasses} (b)). The second new glass frame (large glasses, F3) with a frame around lens (Fig.~\ref{Fig: different glasses} (c)) has a much larger size and weight compared to F1 and F2. To evaluate our system on these new glass frames, three participants from the original study (P1, P5 and P7) agreed to participate in this additional study. The study setups and procedures were exactly the same as the previous study described in Sec.~\ref{Sec: user study}.

We collected 12 sessions of data for each participant testing each glass frame. Since Subsec.~\ref{Subsec: user-adaptive} indicates that 6 sessions of training data is sufficient, we discarded the first 4 sessions for each glass frame and used the last 8 sessions to run a 4-fold cross validation in order to test the tracking performance of our system on different glasses. In this case, we can make sure that participants are familiar with the wearing of all the glass frames and eliminate the impact of some random factors. The evaluation result shows that the small glasses (F2) yielded a similar average performance to the original glasses (F1) (both at $5.3\degree$), while the large glasses (F3) resulted in a relatively poorer average performance (at $6.1\degree$), with a drop in performance of 15\%. One possible reason for the performance difference is that the sensors on the larger glasses were much closer to the skin. Sometimes, the sensors may directly touch the skin, which could block the transmission and reception of signals, as we explained in Subsec.~\ref{Subsec: form factor}.

\subsection{Evaluation on the Final Prototype}
\label{Subsec: Final prototype}
In the previous user studies, we evaluated \gls{name} with various configurations under different scenarios, using the initial prototype that we had developed. The results of these studies helped us confirm the prototype settings and develop an optimized system prototype, which features a more compact form factor as shown in Fig.~\ref{Fig: hardware} (g). In this subsection, our objective was to assess the performance and power consumption of this final prototype.

\subsubsection{Gaze Tracking Accuracy}
To evaluate the final prototype, we recruited 10 participants (four of whom participated in the previous study). The study design was similar to the previous study, except that we only used instruction points as the ground truth acquisition method. Each participant collected eight sessions of data (six sessions for training and two for testing). We reduced the signal strength from the speaker to 20\% of the original setup, as we found that even with 2\% of the original strength, the performance was similar in the pilot study. Hence, this final prototype has significantly lower signal strength and improved environmental sustainability. Additionally, we set the CPU speed of the Teensy 4.1 to 150 MHz in this study (standard speed: 600 MHz) to lower power consumption. With this setting, the system experienced a data loss rate of 0.002\%, and the performance of our system was not affected by this loss, as shown in Tab.~\ref{Tab: attachable prototype}. Apart from the cross-session performance, we also conducted a test of the in-session tracking accuracy in which the training data and testing data were split from the same sessions without remounting the device to show the optimal performance of our system.

\begin{table}[h]
\caption{Gaze Tracking Performance in MGAE with the Final Prototype.}
\label{Tab: attachable prototype}
\scalebox{0.7}{\begin{tabular}{| c | c | c | c | c | c | c | c | c | c | c | c |} 
\hline
 Settings & P1 & P2 & P7 & P10 & P15 & P16 & P17 & P18 & P19 & P20 & Avg\\ [0.5ex] 
 \hline\hline
 Cross-session & $4.4\degree$ & $4.9\degree$ & $5.9\degree$ & $8.0\degree$ & $3.6\degree$ & $3.6\degree$ & $3.9\degree$ & $4.8\degree$ & $4.7\degree$ & $5.7\degree$ & $\textbf{4.9\degree}$\\
 \hline
In-session & $3.9\degree$ & $3.6\degree$ & $4.9\degree$ & $5.3\degree$ & $1.8\degree$ & $2.6\degree$ & $2.6\degree$ & $3.6\degree$ & $3.1\degree$ & $4.9\degree$ & $\textbf{3.6\degree}$\\
 \hline
\end{tabular}}
\end{table}

As shown in Tab.~\ref{Tab: attachable prototype}, the mean gaze angular error (MGAE) is $4.9\degree$ for the cross-session evaluation, which is similar to the previous study. When evaluating the performance of \gls{name} within the same sessions, the accuracy improves to $3.6\degree$. We did not add ear loops to this prototype because the legs of the glasses were wider than the ear loops we had. For most participants, the glasses fit well on their ears, but one participant (P10) reported that the glasses kept sliding down during the study, which may have affected their performance. Based on the questionnaires, no participant reported being able to hear the signal emitted from our system. We also measured the signal level from our system using the NIOSH app. We placed the phone running the app close enough to the speakers in our system and the app gave us an average signal level of 43.1 dB(A). This is below the maximum allowable daily noise recommended by CDC, which is 85 dB(A) over eight hours in the workspace \cite{cdc1998niosh}.

\subsubsection{Power Consumption}
We measured the power consumption of our system with a current ranger \cite{currentranger}. The average current flowing through the system was measured as 88.3 mA @ 3.26 V, which gives us a power consumption of 287.9 mW. This value was tested with all 8 microphones and 2 speakers working, and with the data being written into the SD card. Our system can last up to 38.5 hours with a battery of similar capacity to Tobii Pro Glasses 3 (3400 mAh), while the working time of Tobii Pro Glasses 3 is only 1.75 hours. If applied to non-eye-tracking glasses, like Google Glass, our system can run for 6.4 hours. It is worth noting that these estimates do not include the power consumption of data preprocessing and deep learning inference running on a local server. We measured the power consumption of different components in our system (Tab.~\ref{Tab: power}). Teensy 4.1 has a high base power consumption, while the sensors (speakers and microphones) consume much less power.

\begin{table}[htbp]
\small
\caption{Power Consumption of Different Components on Teensy 4.1. Power of data preprocessing and deep learning inference is NOT included.}
\label{Tab: power}
\scalebox{0.85}{\begin{tabular}{| c | c | c | c |}
\hline
\textbf{Total} & Speakers \& Mics & SD card writing & Other operations\\ [0.5ex] 
 \hline\hline
\textbf{287.9 mW} & 16.4 mW & 72.7 mW & 198.8 mW\\
 \hline
\end{tabular}}
\end{table}

\subsubsection{Usability}
After the user study, we distributed a questionnaire to every participant to ask for feedback on our prototype. First, the participants evaluated the overall comfortableness and the weight of the prototype with a rating from 0 to 5. Across all 10 participants, the average scores they gave to these two aspects are 4.5 (std=0.7) and 4.2 (std=0.8), indicating that \gls{name} is overall comfortable to wear and easy to use. Furthermore, all 10 participants answered "No" to the question "Can you hear the sound emitted from our system?", verifying the inaudibility of the acoustic signals from the \gls{name} system.

\section{Inference on MAX78002}
\label{Sec: MAX78002}
In the previous evaluation, we recorded audio data with Teensy 4.1 first and run the signal processing and deep learning pipeline on a local server offline. To enable predictions of gaze positions in real-time on an MCU, we implemented the whole pipeline on a micro-controller with an ultra-low-power CNN accelerator (MAX78002~\cite{MAX78002}).

\subsection{ML Models}
To achieve this goal, the deep learning models were trained and synthesized in advance, using the ai8x libraries~\cite{ai8x}. We implemented two models with ai8x, which were ResNet-18 (used in the previous study) and MobileNet for comparison. Due to the hardware limit of MAX78002, we modified the models to be compatible with the chip. Specifically, for a Conv2d layer, the kernel size could only be set to 1x1 or 3x3 and the stride is fixed to [1, 1]. In addition, some convolution layers of ResNet-18 were substituted with depthwise separable convolution layers to avoid exceeding the limit of the number of parameters in the model. Furthermore, we quantized the input and the weights of the models with ai8x, which converted them all into 8-bit data format to save memory for storage and increase the speed of inference.

\subsection{Data Preprocessing}
Before the deep learning model, we also need to apply a band-pass filter on the received signals and perform cross-correlation between received signals and transmitted signals to obtain echo profiles as described in Subsubsec.~\ref{SubsubSec: Echo Profiles}. In the implementation on MAX78002, to reduce processing time, we removed the band-pass filter since all the computations are done on the MCU and transmitting private data is no longer a concern.

Then we experimented two different methods to realize the cross-correlation: (1) brute force to calculate echo profiles point by point; (2) the dot product function in the CMSIS-DSP library. Results of standard tests \cite{ai8x-power} revealed that it took the system 178.3 ms and 45.4 ms to compute one echo frame and make one inference with these two methods utilized respectively. Considering that one frame of our audio data comes every 12 ms in our system (600 samples $\div$ 50000 samples/s), this processing time is too long to keep our system running in real-time with an FPS of 83.3 Hz. Finally, we explored method (3) a Conv2d layer (kernel size 1x1) with transmitted signals as the untrained weights and received signals placed along the channel axis of the input. This can increase the speed of echo profile calculation because it uses the CNN accelerator on MAX78002. We compressed the samples used for cross-correlation from 600x600 to 34x34 and the pixels of interest from 60 pixels (20.4 cm) to 30 pixels (10.2 cm) in this case to further decrease the processing time.

With this Conv2d layer added on top of the deep learning model, the model directly takes the raw audio data as input in instances with the size of 64 (34+30 samples) x 26 (frames) x 8 (microphones). This method allows the system to make one inference within 10.3 ms, which is enough for the real-time pipeline with a double-buffer method applied (DMA moves the current frame in one buffer while the CPU processes the previous frame in another buffer).

\subsection{Accuracy and Refresh Rate}
To validate these modifications and compression, we evaluated the in-session performance of different models with different settings using data collected with the final prototype in Subsec.~\ref{Subsec: Final prototype} and showed the results in Tab.~\ref{Tab: ai8x results}.

\begin{table}[h]
\caption{Average In-session Performance across 10 Participants with Different Models and Settings.}
\label{Tab: ai8x results}
\scalebox{0.8}{\begin{tabular}{| c | c | c | c | c |} 
\hline
Models & ML Libraries & Compressed? & Quantized? & MGAE\\ [0.5ex] 
\hline\hline
\multirow{4}{*}{ResNet-18} & pytorch & \xmark & \xmark & $3.6\degree$\\
\cline{2-5}
& ai8x & \xmark & \xmark  & $4.0\degree$\\
\cline{2-5}
& ai8x & \cmark & \xmark & $4.0\degree$\\
\cline{2-5}
& ai8x & \cmark & \cmark & $4.2\degree$\\
\hline
\multirow{2}{*}{MobileNet} & ai8x & \cmark & \xmark & $4.2\degree$\\
\cline{2-5}
& ai8x & \cmark & \cmark & $4.3\degree$\\
\hline
\end{tabular}}
\end{table}

As shown in the table, the same model trained with ai8x is slightly worse than that trained with PyTorch given the constraints of the convolution layers discussed above. Compressing the size of input data does not affect the accuracy. While MobileNet yields comparable accuracy to ResNet-18, both models suffer a slight performance drop after quantization since the precision of data is decreased.

Given the limitation of the I2S interfaces on MAX78002, to test our system in a more realistic condition, we still use Teensy 4.1 to control the speakers and microphones and transfer the received audio data to MAX78002 via the serial port. To accelerate the transmission speed, only the samples that are used for processing on MAX78002 are transferred. This generates a steady stream of audio data to MAX78002. In future, we will explore connecting microphones directly to MAX78002 using multi-channel audio protocols such as Time-division Multiplexing (TDM). Evaluation results showed that for ResNet-18 and MobileNet, MAX78002 spent 124.1 ms and 41.6 ms respectively loading the weights of the model. This is a one-time effort and can be done before running the real-time pipeline so it did not impact the refresh rate. Then it took 12 ms to load one instance and make an inference based on it in real-time for both ResNet-18 and MobileNet, giving a refresh rate of 83.3 Hz.

\subsection{Power Consumption}
We measured the power consumption of the MAX78002 evaluation kit while it made inferences. Tab.~\ref{Tab: power max78002} demonstrates that MAX78002 consumes 96.9 mW and 86.0 mW respectively when making inferences with ResNet-18 and MobileNet at 83.3 Hz. The refresh rate can be reduced to 30 Hz to save power, which is enough for many applications. In this case, the power becomes 79.0 mW and 75.7 mW respectively.

If we can use MAX78002 to directly control speakers and microphones in future, we will be able to optimize the power efficiency and keep the overall power consumption of our real-time system around 95.4 mW, i.e., 79.0 mW (MAX78002 with ResNet-18 running at 30 HZ) + 16.4 mW (2 speakers and 8 microphones). One should keep in mind that this is just an estimate of the power of this real-time system and the power consumption of MAX78002 might increase if it does need to control the sensors but we do not expect it to be very high because the current power of MAX78002 already includes that of the CPU and the CNN accelerator running at full speed.

\begin{table}[htbp]
\small
\caption{Power Consumption of MAX78002 with Different Models Running.}
\label{Tab: power max78002}
\begin{tabular}{| c | c | c | c | c |} 
\hline
 Models & \multicolumn{2}{|c|}{ResNet-18} &  \multicolumn{2}{|c|}{MobileNet}\\ [0.5ex] 
 \hline\hline
 FPS (Hz) & 83.3 & 30 & 83.3 & 30\\ [0.5ex]
 \hline
 Power (mW) & 96.9 & 79.0 & 86.0 & 75.7\\
 \hline
\end{tabular}
\end{table}

\section{Discussion}
\label{Sec: Discussion}

\subsection{Evaluating Simpler Regression Models}
We adopted two traditional regression models, which are linear regression (LR) and gradient boosted regression trees (GBRT), to predict gaze positions using the data collected in Subsec.~\ref{Subsec: Final prototype} and the results showed that the average in-session tracking accuracy for these two models across 10 participants are 11.6$\degree$ and 6.8$\degree$ respectively. Compared to the results in Tab.~\ref{Tab: attachable prototype}, the traditional regression models output much worse accuracies than ResNet-18 (3.6$\degree$). We conducted an analysis of the impurity-based feature importance with GBRT, comparing the features in different channels of microphones in Tab.~\ref{Tab: feature importance}. It turns out that the channels receiving signals from 18-21 kHz (S1) are generally more important than channels receiving signals from 21.5-24.5 kHz (S2). Furthermore, the microphones that are closer to the inner corners of the eyes (M1, M4, M5 M8) are more important than those closer to the tails of the eyes (M2, M3, M6, M7).

\begin{table}[h]
\caption{Feature Importance Analysis for Different Microphones using GBRT (Scaled to 0-100).}
\label{Tab: feature importance}
\scalebox{0.7}{\begin{tabular}{| c | c | c | c | c | c | c | c | c |} 
\hline
Microphones & M1 & M2 & M3 & M4 & M5 & M6 & M7 & M8\\ [0.5ex] 
\hline\hline
Importance (S1/S2) & 100/27 & 57/23 & 33/12 & 96/29 & 66/81 & 42/17 & 28/15 & 85/79\\
 \hline
\end{tabular}}
\end{table}

\subsection{Impact of Real-world Factors}

\subsubsection{Head Movements}
In the user study, we did not use a chin rest to fix the participants’ head so they could turn their head freely. However, we believe that how head movements affect the system performance should be evaluated in more details in the future.

\subsubsection{Near- and Far-sighted}
In the user study, we collected participants’ degrees of myopia in the questionnaires, which showed no connection to the gaze tracking performance.

\subsubsection{User Speaking}
One researcher evaluated our system when keeping silent and keeping talking to himself. The gaze tracking performance of the silent sessions and the talking sessions is the same at $3.9\degree$.

\subsection{Potential Applications}

The goal of this paper is to demonstrate the feasibility of our new acoustic-based gaze tracking system on glasses. While our eye tracking accuracy of $4.9\degree$ is comparable to some webcam-based methods, it is lower than commercial eye trackers ($1.9\degree$ in our study). Therefore, our system may not be immediately applicable to some applications requiring highly precise eye tracking. However, our system can still be used in many applications, such as interaction with interface elements like buttons in AR, that do not require very high accuracy eye trackers. 

Our system can also be potentially used in tracking irregular eye movements, enabling healthcare applications for monitoring users' health conditions in everyday life. This requires monitoring the gaze movements throughout the day for analysis in everyday life, instead of just tracking their accurate gaze positions for a few hours in a controlled settings. The low-power and lightweight features of our \gls{name} system make it a good candidate solution to enabling a variety of applications that camera-based eye trackers cannot realize, by continuously understanding user gaze movements in the wild for extended periods. Furthermore, our system can alleviate the privacy concern from users as compared to camera-based methods. 

\subsection{Limitation and Future Work}
\subsubsection{Improving the Performance} 
There is room for further improvements of the performance of our system. For instance, we can apply calibration process on the output of the system to further enhance performance. We experimented with affine transformation and projective transformation to transform the output but they did not immediately improve the performance. According to the analysis of the error distribution of the eye tracking results, we believe this is because the error distribution is not linear in our system so we need to explore more non-linear transformation methods to improve the performance.

\subsubsection{Calibration Process for Fine-tuning}
Our system currently requires a 15-second calibration process before each session to fine-tune the model, which may be inconvenient for users. However, Subsec. \ref{Subsec: user-dependent} shows that the tracking accuracy without fine-tuning is still acceptable, at $5.9\degree$, compared to the accuracy achieved with fine-tuning ($4.9\degree$). 

\subsubsection{Reducing Training Effort}
Subsec.~\ref{Subsec: user-adaptive} suggests that \gls{name} achieves satisfactory performance on new users with approximately 20 minutes of training data using the user-adaptive model. This training effort can be further reduced by constructing a larger and more diverse dataset from much more participants to train the based model. Moreover, data augmentation methods, such as including simulation data to train the model, can be explored as well.

\subsubsection{Towards a More Integrated System}
In Sec.~\ref{Sec: MAX78002}, we still used Teensy 4.1 to control the speakers and microphones, and transfer audio data to the MCU MAX78002. In future, we plan to further customize our own PCBs for MAX78002 to allow it to directly control speakers and microphones. We believe that the power consumption of our real-time system can be further reduced in this case because Teensy 4.1 with a high base power can be removed. Furthermore, we do not expect that the power consumption of MAX78002 will be significantly increased since the on-board CPU and CNN accelerator of MAX78002 were already operating at maximum speed in our current evaluation. With a solid system implementation, we plan to carry out an extensive evaluation of this more integrated system in future work to validate our speculation.
\section{Conclusion}
\label{Sec: Conclusion}

In this paper, we present the first acoustic-based eye tracking glasses capable of continuous gaze tracking. The study involving 20 participants confirms that our system can accurately track gaze points continuously, achieving an accuracy of $3.6\degree$ within the same session and $4.9\degree$ across different sessions. When compared to commercial camera-based eye tracking glasses such as Tobii Pro Glasses 3, our system reduces power consumption by 95\%. A real-time pipeline is implemented on MAX78002 to make inferences with a power signature of 95.4 mW at 30 Hz.
\begin{acks}
This work is supported by National Science Foundation (NSF) under Grant No. 2239569, NSF's Innovation Corps (I-Corps) under Grant No. 2346817, the IGNITE Innovation Acceleration Program, and the Ann S. Bowers College of Computing and Information Science at Cornell University. The authors would like to thank Prof. Susan Fussell and her lab for sharing the access to the commercial eye tracker Tobii Pro Fusion with us. We also appreciate the help of a Cornell student, Vipin Gunda, on the development of the pipeline on MAX78002.
\end{acks}

\bibliographystyle{ACM-Reference-Format}
\bibliography{sample-base}


\end{document}